# Inelastic neutron scattering: resonances and fluctuations

F. W. K. Firk

*The Henry Koerner Center for Emeritus Faculty,*

*Yale University, New Haven CT 06520*

**Abstract.** The R-function theory of Thomas is used in a study of the inelastic scattering of neutrons from states of the same spin and parity to a definite state. The onset of fluctuations, the effects of randomness of the phases of the reduced width amplitudes for elastic and inelastic scattering, and the effects on fluctuations of statistical variations in the distribution of the neutron widths, are considered in detail. In the region of strongly overlapping resonances, the calculated cross sections exhibit narrow structures that are highly sensitive to the phases of the amplitudes involved. In the resolved resonance region, very sharp interference effects are found to occur between certain adjacent resonances with appropriate relative widths and relative phases. The Thomas approach is not limited to the study of nuclear reactions. Studies of conductance fluctuations associated with quantum dots in semiconductors at low temperature, and fluctuations observed in electron-atom interactions that involve many channels and many overlapping resonances, require analyses of the kind reported here. In all cases, it is necessary to achieve the highest possible energy resolution to observe the sharp fluctuations found in the present analysis.

NUCLEAR REACTIONS. Thomas' R-function theory is used to model neutron inelastic scattering to a definite excited state. Sharp fluctuations in regions where $<\Gamma>/<D> >> 1$ for states of the same spin and parity, and strong interference effects in the region of resolved resonance, are predicted.

# 1. Introduction

The problems associated with using the general **R**-matrix theory of Wigner and Eisenbud (1948) to analyze observed cross sections in many-level, many-channel nuclear reactions, electron-atom/molecule interactions, and conductance fluctuations associated with quantum dots in semiconductors, are well known (Lane and Thomas 1958, Vogt 1958, 1960, Burke and Berrington 1993, Alhassid 2000 and Weidenmueller 2002). These problems are particularly severe in energy regions where many levels (of a given spin and parity) overlap. In order to obtain cross sections from an **R**-matrix, and its associated collision matrix, it is necessary to carry out challenging matrix inversions. The matrices involved are invariably of large dimension, and the values of the required matrix elements are known only for one or two channels. In the mid-1950's, a major development in the practical use of **R**-matrix theory was made by Thomas (1955); his work is outlined in Chapter 2. At that time, the results of experiments in the field of low-energy neutron resonance spectroscopy were not sufficiently precise to answer, definitively, questions concerning the spacing distribution of adjacent resonances, and the distribution of the (elastic) neutron widths of the resonances. In 1956, Wigner (1957, 1958) conjectured that the spacing distribution is the same as that associated with the spacing distribution of the eigenvalues of random matrices that represent the intractable nuclear Hamiltonian, and Porter and Thomas (1956) proposed that the distribution of neutron widths is a chi-squared distribution with one degree of freedom (see also, the earlier work of Brink (1955)).

In the years 1956 – 1960, improved experiments showed that both distributions are valid (Harvey and Hughes 1958, Rosen *et al* 1960, and Firk, Lynn and Moxon 1960). Lane and Lynn (1957) applied the new statistical ideas to the analyses of cross sections in the region of resolved neutron resonances. They discussed, for example, the effects of the distributions on the average cross sections for radiative capture. Egelstaff (1958) introduced an auto-correlation function in his analysis of the fluctuations in the neutron strength functions of heavy elements observed in measurements of the neutron total cross sections of several heavy nuclei. Egelstaff's analysis pre-dated Ericson's introduction of auto-correlation functions in the analysis of fluctuations observed in nuclear reaction cross sections at energies where $\langle\Gamma\rangle/\langle D\rangle > 1$ (Ericson (1960, 1963)). The importance of the role played by the phases of the amplitudes was first demonstrated in detailed analyses of neutron-induced fission cross sections in the resonance region (Vogt 1958, Reich and Moore 1958). All aspects of the theory of neutron resonance reactions, and the relevant experiments, are discussed in the standard work of Lynn (1968).

In Ericson's method, explicit forms of the spacing and width distributions are not invoked, and questions concerned with possible variations in the phases of the interfering amplitudes are not addressed; it is a method that involves *average*s. Later, Moldauer (1964a, b) included, also in an average way, the effects of width fluctuations in an analysis of overlapping resonance at high excitation energies, and showed that the traditional method of statistical analysis was thereby improved.

In this paper, the statistical distributions are included in a microscopic study of inelastic neutron scattering, to a definite state; *no averaging is carried out*. It is therefore possible to study not only the development of Ericson and Moldauer fluctuations as the ratio <Γ>/<D> increases from less than 1 to greater than 1, but also the effects of *random variations* in the *statistical distributions*, and in the *phases* of the *reduced width amplitudes*. Sharp fluctuations are found in the calculated cross sections; they persist for values <Γ>/<D> >> 1, and differ in character from either Ericson or Moldauer fluctuations.

In the resonance region, strong interference effects are expected between nearby resonances that have favorable widths, and relative phases of their interfering amplitudes.

## 2. Theory

*2.1. Inelastic scattering to a definite final state: defining the problem*

The general problem concerns the inelastic scattering of a particle from highly excited states of a many-body system to a definite final state. The inelastic scattering is accompanied by elastic scattering, and by transitions to many alternative channels. The specific problem concerns the inelastic scattering of a neutron by an even-even nucleus to a definite state. Elastic scattering is included explicitly, and inelastic scattering to all other allowed states, and radiative capture, are treated in an average way. Each state, $\lambda$, with a spin and parity $J^\pi$ is characterized by an energy $E_\lambda$ and a total width $\Gamma_\lambda$. An incident neutron interacts with a heavy nucleus to form a state that decays into many different channels. The two main channels are inelastic scattering to a definite state (width $\Gamma_{n'}$), and elastic scattering (width $\Gamma_n$). All the subsidiary channels involve inelastic scattering (width $\Gamma_{n''i}$) and radiative capture (width $\Gamma_{\gamma i}$). (At higher energies, it is necessary to include possible proton decay channels in the eliminated width). The cross section for inelastic scattering to the definite state is calculated using the Thomas (1955) approximation to the general **R**-matrix theory of Wigner and Eisenbud.

*2.2. The Thomas approximation*

The **R**-matrix has the energy-dependent form (Lane and Thomas 1958)

$$\mathbf{R}(E) = \sum_\lambda \gamma_{\lambda c'} \gamma_{\lambda c''}/(E_\lambda - E)$$

where the sum is over *all* levels $\lambda$ of energy $E_\lambda$, and the $\gamma_{\lambda c}$'s are the reduced width amplitudes associated with the channels $c'$, $c''$. If the signs of the amplitudes are sufficiently random to ensure that the non-diagonal elements of **R** are small compared with the diagonal elements, Thomas showed that the resulting collision matrix **U** can be written

$$\mathbf{U} = \sum_\lambda (a_{\lambda c'} a_{\lambda c''})/(E_\lambda - E - i\Gamma_\lambda/2), \ c' \neq c'',$$

even for *overlapping states*. The amplitudes, $a_{\lambda c}$'s are given by

$$a_{\lambda c} = \gamma_{\lambda c}\sqrt{(2P_c)},$$

where $P_c$ is the penetration factor, and the width is

$$\Gamma_{\lambda c} = a_{\lambda c}^2.$$

The width that occurs in the denominator is

$$\Gamma_\lambda = \sum_c \Gamma_{\lambda c},$$

in which

$$\Gamma_{\lambda c} = |a_{\lambda c}/(1 - iP_c(R_c^\infty + i\pi\rho<\gamma_{\lambda c}>^2))|^2,$$

$R^\infty$ is the effect of all states outside the range of interest, $\rho$ is the density of states, and $<...>$ denoted the average value in the region of E.

The cross section for the reaction c' –> c" is

$$\sigma_{c'c''} = (\pi/k_{c'})^2 \sum_{c',c''} |U_{c',c''}|^2$$

where $k_{c'}$ is the wave number of the relative motion of the two particles in the incident channel. (The spin weighting factor has been put equal to unity).

In the present case that involves two main channels and many subsidiary channels, Thomas showed that a *reduced* form of the **R**-matrix is valid; it is

$$R(E) = \sum_\lambda (\gamma_{\lambda c'} \gamma_{\lambda c''})/(E_\lambda - E - i\Gamma_\lambda^e/2), \text{ the } R\text{-function},$$

where $\Gamma_\lambda^e$ is a suitable average of the widths of all the subsidiary, or *eliminated* channels. Here, the eliminated channels are all channels except the incident channel, and the inelastic scattering channel to the definite state. *The reduced **R**-matrix is valid if the means of the partial widths for the eliminated channels are less than the spacings, and their amplitudes are random in sign.* It is seen that the reduced **R**-matrix can be obtained from the traditional **R**-matrix by evaluating it at the complex energy $E = E + i\Gamma_\lambda^e/2$. (The level shifts have been put equal to zero).

Reich and Moore (1958) developed the Thomas R-function theory and applied their method to an analysis of cross sections for neutron-induced fission in the low-energy resonance region.

In a detailed analysis of fluctuations in the continuum Ericson (1963) discussed the fundamental scattering matrix on which the model was based; it had been shown by Feshbach (1962) to be *valid for overlapping states as an average concept.* The importance of Feshbach's scattering matrix in the theory of Ericson fluctuations cannot be overstated.

*2.3. Cross section for inelastic neutron scattering to a definite state*

The cross section for the inelastic scattering of a neutron to a definite state in the presence of elastic, all other inelastic channels, and radiative capture is calculated using the Thomas R-function. The calculations are restricted to s-wave scattering in the incident and exit channels; the restriction is expedient from a calculational point of view, and does not represent a fundamental limitation of the method. The Thomas approach leads to the following expression for the inelastic scattering cross section:

$$\sigma_{n,n'}(k_n^2/4\pi) = \left| \frac{\sum_\lambda (\Gamma_{\lambda n}/2)^{1/2}(\Gamma_{\lambda n'}/2)^{1/2}/f_\lambda(E)}{[1 - i\sum_\lambda(\Gamma_{\lambda n}/2)/f_\lambda(E)][1 - i\sum_\lambda(\Gamma_{\lambda n'}/2)/f_\lambda(E)] + [\sum_\lambda(\Gamma_{\lambda n}/2)^{1/2}(\Gamma_{\lambda n'}/2)^{1/2}/f_\lambda(E)]^2} \right|^2$$

where

$$f_\lambda(E) = E_\lambda - E - i\Gamma_\lambda^e/2,$$

and the sums are over *all* states λ of the same spin and parity. The width of the eliminated channels is

$$\Gamma_\lambda^e = \sum_{n''}\Gamma_{\lambda n''} + \sum_i \Gamma_{\lambda\gamma i} = \Gamma_\lambda - (\Gamma_{\lambda n} + \Gamma_{\lambda n'}),$$

where $\Gamma_\lambda$ is the total width, $\Gamma_{\lambda n}$ is the elastic scattering width, $\Gamma_{\lambda n'}$ is the inelastic scattering width to the definite state, $\Gamma_{\lambda n''}$ is an inelastic scattering width to an eliminated state, and $\Gamma_{\lambda\gamma i}$ is a partial radiation width to an eliminated state. The neutron wave number associated with the incident channel is $k_n$, and the spin weighting factor is unity for s-wave scattering from an even-even nucleus.

In the expression for $\sigma_{nn'}$, interference terms of the form

$$\sum_\lambda [(+/-)\Gamma_{\lambda n}^{1/2}\Gamma_{\lambda n'}^{1/2}(E_\lambda - E)]/[(E_\lambda - E)^2 + (\Gamma_\lambda^e/2)^2]$$

and

$$\sum_\lambda [(+/-)\Gamma_{\lambda n}^{1/2}\Gamma_{\lambda n'}^{1/2}(\Gamma_\lambda^e/2)]/(E_\lambda - E)^2 + (\Gamma_\lambda^e/2)^2]$$

are sensitive to the effects of the (random) phases associated with the primary elastic and inelastic scattering amplitudes.

The cross section $\sigma_{nn'}$ is calculated for up to 1000 interfering states at up to 10000 energies. The spacing between pairs of adjacent states is chosen randomly from a Wigner distribution, and the reduced width amplitudes for the n and n' channels are chosen randomly from Porter-Thomas distributions. Before the publication of the work of Porter and Thomas, Brink (1955) had proposed that the reduced width amplitudes for a single channel process at high excitation energy should have random signs associated with a normal distribution. In the present case, the signs of both the elastic and inelastic scattering amplitudes that appear in the expression for $\sigma_{nn'}$ are chosen randomly.

**3. Results**

*3.1. Resolved resonance region: threshold inelastic scattering*

Experiments are now being proposed to measure inelastic neutron scattering cross sections to very low-lying states, typically 10 keV above the ground state. These studies will involve high resolution "threshold inelastic scattering". The forms of the interfering components of the inelastic neutron cross section given by the Thomas theory lead to sharp interference patterns in nearby resonances that have the appropriate relative widths and phases.

3.1.1. Interference patterns in the region of resolved resonances

The calculated inelastic neutron scattering cross section to a low-lying state at 10 keV is shown in Figure 1. The parameters are i) average spacing $\langle D \rangle$ = 100 eV, and ii) $\langle \Gamma \rangle / \langle D \rangle$ = 0.2. The spacings are chosen randomly from a Wigner distribution, and the elastic and inelastic neutron widths are chosen randomly from Porter-Thomas distributions. The widths of the eliminated channels, $\Gamma^e$ = 0.1 eV, are constant. An inspection of the resonant line shapes shows that, in favorable cases, sharp interference effects are found to occur.

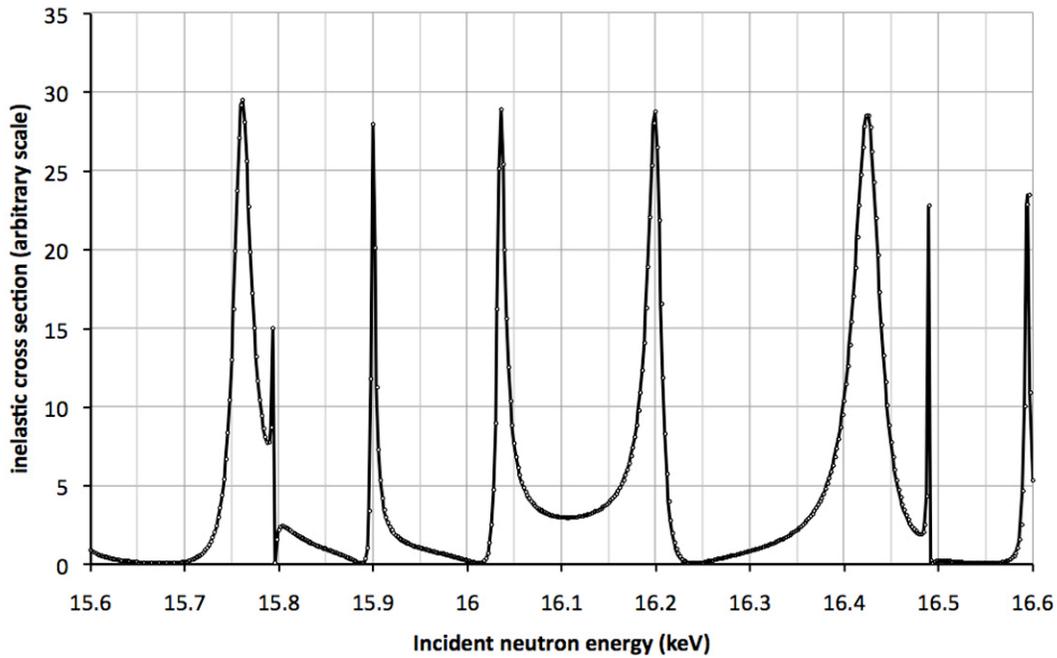

 Figure 1.  The calculated inelastic neutron scattering cross section to a state at 10 keV.  The pairs of resonances at 15760 and 15900 eV, and 16425 and 16490 eV, exhibit strong interference effects that depend on the relative widths and the phases associated with the (random) interfering amplitudes of the elastic and inelastic channels.

To observe these characteristic interference forms it is necessary to achieve high resolution.  The assignment of the (correct) phases to the reduced width amplitudes for the elastic and inelastic channels will need to be done in an iterative fitting process.

*3.2.  Onset of fluctuations*

The inelastic neutron scattering cross section to a state at 1 MeV is calculated for two values of the strength function, $<\Gamma>/<D> = 3$, and 10; the results are shown in Figure 2.

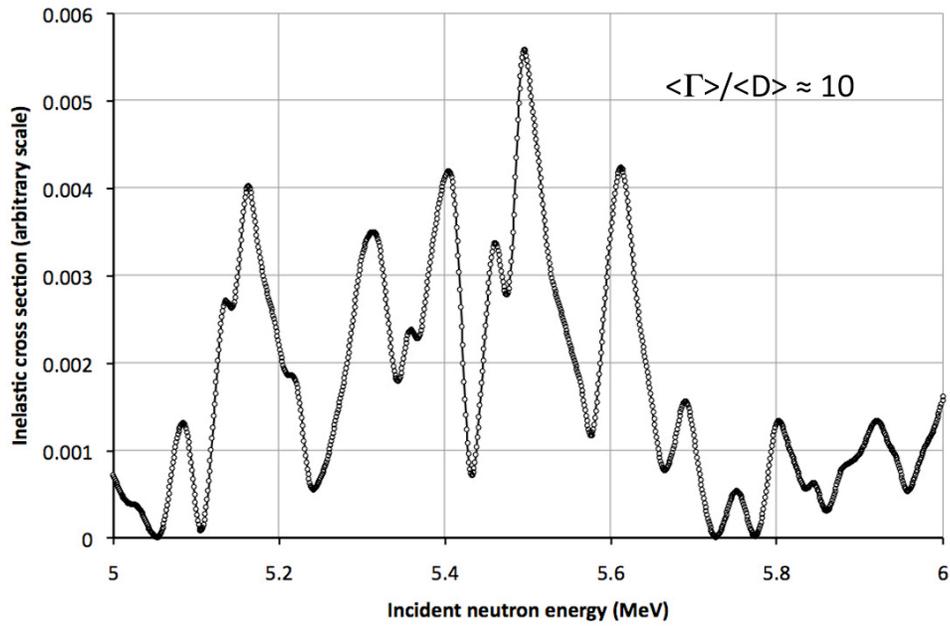

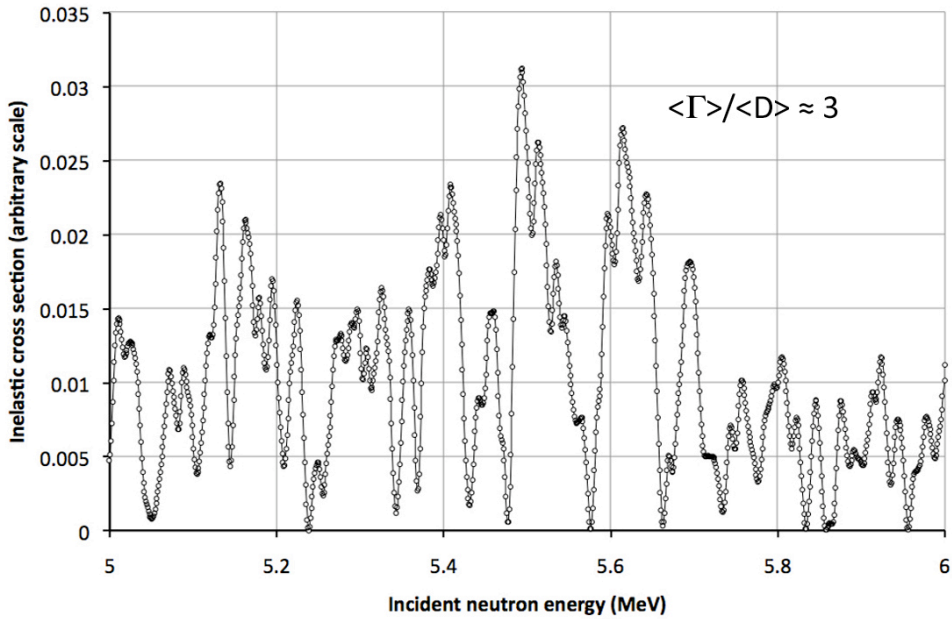

Figure 2. The calculated inelastic neutron scattering cross section to a state at 1 MeV for two values of the strength function: $\langle\Gamma\rangle/\langle D\rangle = 3$ and 10. The fine-structure resonance energies are the same in the two cases.

The parameters are i) an average spacing, <D> = 5 keV, for the underlying fine-structure resonances. The spacings are chosen randomly from the same Wigner distribution, using the same set of random numbers, ii) neutron widths for elastic and inelastic scattering chosen randomly from Porter–Thomas distributions. For both values of the strength function, a constant, average value for the sum of the eliminated widths is assumed. The onset of fluctuations is clearly seen; they continue to develop until <Γ>/<D> ≈ 10.

*3.3. Random phase effects*

The calculated inelastic neutron scattering cross sections to a state at 1 MeV, for two sets of random phases of the interfering amplitudes, in the energy range 5.0 – 6.0 MeV, are shown in Figure 3a.

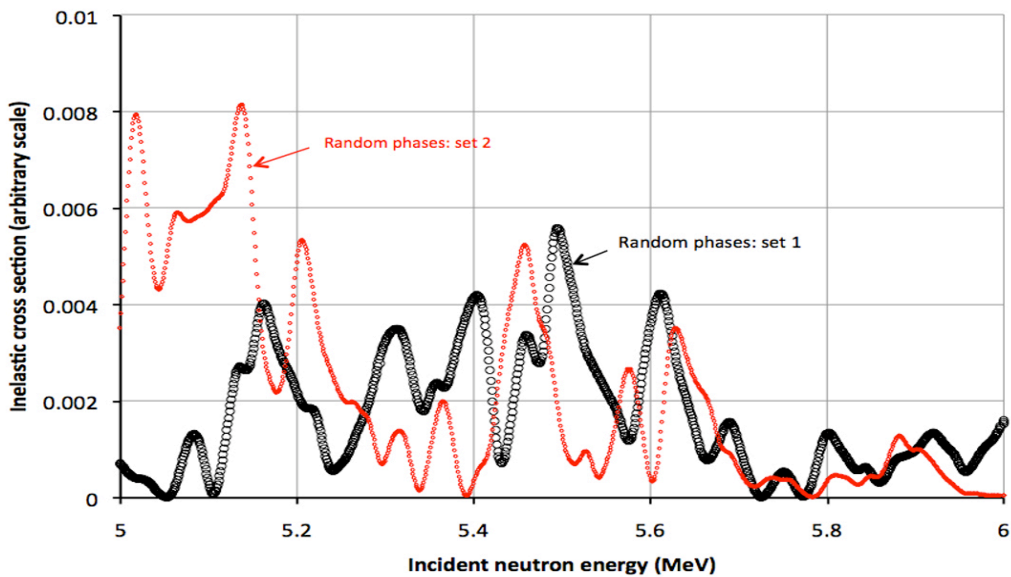

Figure 3a. Calculated inelastic neutron scattering cross section to an excited state at 1 MeV for two different sets of random phases. All other parameters are unchanged.

The common value of the strength function is $<\Gamma>/<D> = 10$. These results illustrate the *strong dependence of the fluctuating cross section on the choice of phases* for the amplitudes of the two channels, n and n'. The "phase fluctuations" are not correlated with the structure of the original, discrete resonances. The unknown phases contribute to the systematic error in any measurement of the spacing distribution of fluctuations made over a limited energy range. The average spacing of the well-defined fluctuations for set 1 is about 90 keV. This value is consistent with that predicted in the theory of Brink and Stephen (Brink and Stephen, 1962), namely $<D>_{fluct} = 2<\Gamma> \approx 100$ keV ($<D> = 5$ keV and $<\Gamma>/<D> = 10$). To illustrate the importance of the *minima* in the pattern of fluctuations it is necessary to plot the cross sections on a logarithmic scale, as shown in Figure 3b.

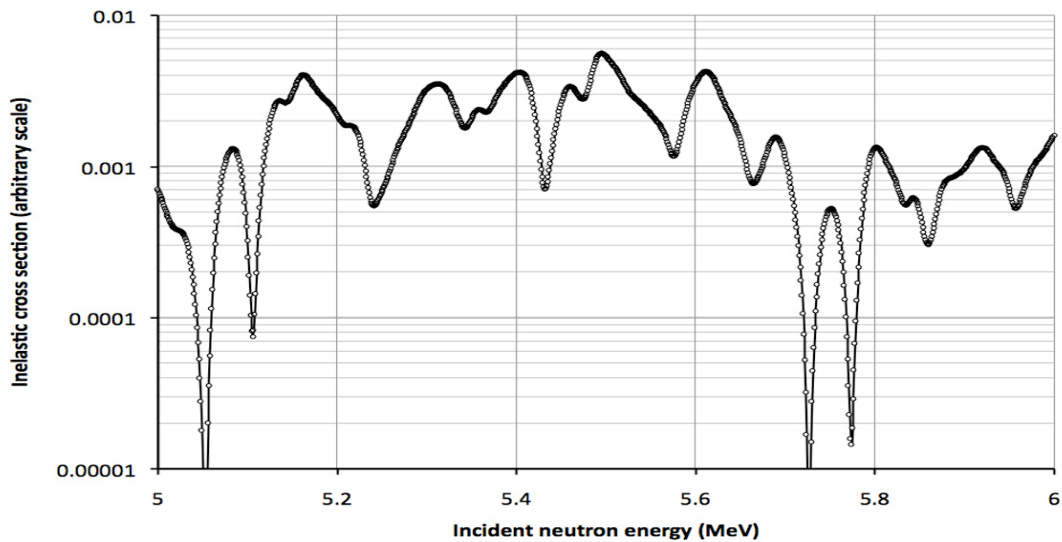

Figure 3b. A log-linear plot of the inelastic neutron scattering to a state at 1 MeV for set 1 of the random phases shown in Figure 3a.

*It is seen that fluctuations are characterized not only by well-defined maxima but also by sharp minima.*

The need for both excellent energy resolution and signal-to-noise ratio is clear from the nature of the calculated sharp variations.

*3.3 Statistical distribution effects*

It has been known since the 1950's that, in any limited energy region, resolved resonances with neutron widths two-to-three times larger than the average value have a major influence on the locally measured strength function. It is therefore important to study the influence on the pattern of fluctuations in the continuum of those underlying resonances with the largest neutron widths. Moldauer (1964a) discussed the effects of varying strength functions on fluctuations; he treated the problem from the point of view of the average properties of the Porter-Thomas distribution function. Here, the problem is addressed by studying changes in the neutron widths of individual resonances.

3.3.1. Variations in the Porter-Thomas distribution of neutron widths

The calculated inelastic neutron scattering cross section to a state at 1 MeV, for standard and modified Porter-Thomas distributions of the inelastic neutron widths, are shown in Figure 4. The incident neutron energy range is 5 – 6 MeV, and the strength function is $\langle\Gamma\rangle/\langle D\rangle \approx 10$.

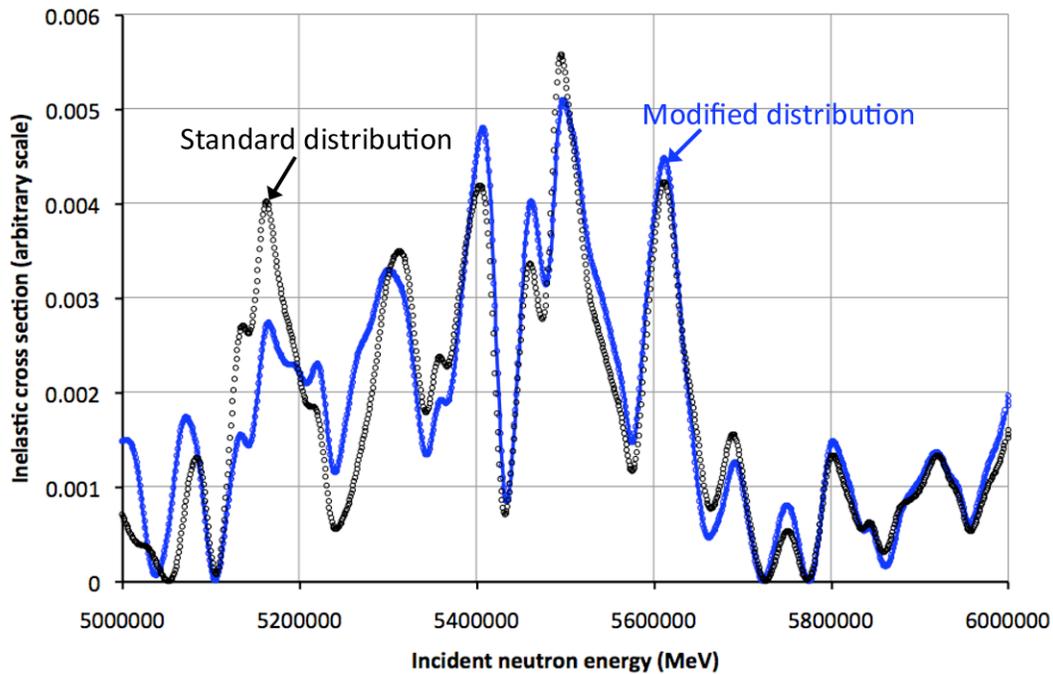

Figure 4. The calculated cross section for inelastic scattering to a state at 1 MeV for i) standard Porter-Thomas distributions for the elastic and inelastic neutron widths and ii) a modified Porter-Thomas distribution for the inelastic widths. In case ii), the inelastic neutron widths of seven of the strongest resonances are reduced to the average value.

In the modified form, seven of the two hundred resonances at energies between 5 and 6 MeV with inelastic neutron widths more than twice the average, have their inelastic neutron widths reduced to the average value. This procedure results in a lowering of the strength function by about 4%. The variation in the small number of large widths has a measurable effect on the fluctuation pattern, most noticeably at energies in the ranges 5 – 5.2, 5.4 – 5.5, and 5.6 – 5.7 MeV. Such variations in the Porter-Thomas distributions of neutron widths are always present in limited energy ranges, and therefore they contribute to the error assigned in any fluctuation analysis. This conclusion, based on a *microscopic* study of fluctuations is in agreement with the original Moldauer (1963) theory, based on *average* properties of the distribution

## 4. Conclusions

A microscopic study of the energy dependence of the inelastic neutron scattering cross section to a definite state has been made using Thomas' R-function theory. The cross sections are studied in the resonance region, and in the continuum. In the resonance region, strong interference effects occur in adjacent pairs of resonances that possess favorable widths and phases associated with the amplitudes for elastic and inelastic scattering. Just above the threshold for inelastic scattering to the first excited state, the eliminated width is the total radiation width. In the continuum, the inelastic neutron scattering cross section to an excited state is shown to depend strongly on random variations in both the phases of the interfering amplitudes and the number of neutron widths that exceed the local average by about a factor of two. These variations are unknowable, and therefore they contribute to the overall uncertainty in any result obtained on the density and magnitude of fluctuations studied in a given range of energy. Furthermore, sharp minima are found to be as characteristic of fluctuation phenomena as the traditional maxima. The method used in the present study is not limited to the field of nuclear reactions; it is applicable in any many-body, many-channel system where inelastic scattering takes place.

## References

Alhassid Y 2000 *Rev. Mod. Phys*. **72** 895

Brink D M 1955 D Phil Thesis, Oxford, unpublished

Burke P G and Berrington K A 1993 *Atomic and Molecular Processes, an R-matrix Approach* (Bristol: Inst. of Phys. Publishing)

Egelstaff P A 1958 *J. Nucl Energy* **7** 35

Ericson T E O 1960 *Phys. Rev. Lett*. **5** 430

——— 1963 *Ann. Phys*. (N.Y.) **23** 390

Feshbach H 1962 *Ann. Phys*. (N.Y.) **19**, 287

Firk F W K, Lynn J E and Moxon M C 1960 *Proc. Int. Conf. Nucl. Struct. (Kingston)* eds D A Bromley and E Vogt (Toronto: Univ. of Toronto Press) p 757

Harvey J A and Hughes D J 1958 *Phys. Rev*. **109** 471

Jalabert Rodolfo A, Stone A Douglas and Alhassid Y 1992 *Phys. Rev. Lett*. **23** 3468

Lane A M and Lynn J E 1957 *Proc Phys. Soc (London)* **A70** 557

Lane A M and Thomas R G 1958 *Rev. Mod. Phys.* **30** 257